# Physics of Wound Healing I: Energy Considerations

S. Peter Apell[1], Department of Applied Physics and Gothenburg Physics Centre,
Chalmers University of Technology, SE-412 96, Göteborg, Sweden

Michael Neidrauer and Elisabeth S. Papazoglou^, School of Biomedical Engineering, Science and Health Systems, Drexel University, Philadelphia, PA 19104 USA

and

Vincent Pizziconi, Harrington Department of Bioengineering
Arizona State University, Tempe, AZ 85287 USA

**Abstract.** Wound healing is a complex process with many components and interrelated processes on a microscopic level. This paper addresses a macroscopic view on wound healing based on an energy conservation argument coupled with a general scaling of the metabolic rate with body mass *M* as $M^\gamma$ where $0 < \gamma < 1$. Our three main findings are 1) the wound healing rate peaks at a value determined by $\gamma$ alone, suggesting a concept of *wound acceleration* to monitor the status of a wound. 2) We find that the *time-scale* for wound healing is a factor $1/(1-\gamma)$ longer than the average internal timescale for producing new material filling the wound cavity in corresondence with that it usually takes weeks rather than days to heal a wound. 3) The model gives a prediction for the maximum wound mass which can be generated in terms of measurable quantities related to wound status. We compare our model predictions to experimental results for a range of different wound conditions (healthy, lean, diabetic and obsese rats) in order to delineate the most important factors for a positive wound development trajectory. On this general level our model has the potential of yielding insights both into the question of local metabolic rates as well as possible diagnostic and therapeutic aspects.

---

[1] corresponding author: apell@chalmers.se

## Physical aspects of wound healing

No doubt wounds have a major impact on living objects. Wounds inflicted in surgical procedures, going to the dentist or by random accidents probably amount to the order of $10^{10}$ per year for the planet. This corresponds to massive sufferings and enormous health care resources which have to be allocated. Not to mention wounds created just because we would like to look differently or temporarily defeat the actions of gravity.

Probably the first cells around would have had to have some mechanism for self-repair when the cell membrane was ruptured [Levine 2009]. In the same way our bodies and our internal organs are surrounded by a membrane; the epithelial tissue with the possibility of healing itself when breached. In this sense the skin, thin but covering some $2\,m^2$, is of interest not only from a physiological point of view but also from a more fundamental angle in being the outermost barrier between ourself and the world around us. Plants also have a wound healing response in the way of chemical defense or simply growing faster (Bloch, 1941, 1952 & Hudler 1984). From the point of harvest damage plant wounds amount to enormous losses in possible resources for a hungry planet (Walter 1983 & 1990) while at the same time being a potential arena for improving our own physiological process of healing a wound.

We have reasons to believe that there is a generic system response at hand in all living systems which connects wound healing to more fundamental questions coupled to the just-in-time energetics, building new matter before knowing the final outcome and the orchestrating role of the immune system. These and other questions are adressed from a physical point of view in this paper being the first, in a series of three. We will consider in order: I) the role of energetics for the wound healing rate (this paper), II) the influence of size, perimeter and other geometrical effects as well as III) the influence of endogenous or exogenous electric fields on wound healing. Even though our treatment is mainly for mammals we do consider our treatment to be a more general one from a generic stand-point. In short we develop an injury model accounting for the wound repair system based on the most important physical factors for living systems healing a breach in their outer protective layers.

The interested reader might of course advocate that wound healing is such a complicated process and depends on so many factors questioning what could possibly be achieved with a physical model. Our rationale is to see what aspect can actually be understood from the shear distribution of energy in the system, using a model derived from first principles without *ad hoc* assumptions and with actual parameter values which can be compared or extracted from experiments. These parameters should then reflect the different ways in which internal and external factors influences the wound healing process. Viewing the long list of possible factors underscores the importance in finding if there are aspects of wound healing where these details do not matter and that there is a more general system level at work. We can divide these factors into three categories. The first is simple physical factors for which we should be able to find a direct linking to our model: pressure (whether positive or negative), pH (Gethin 2007, Schneider 2007), oxygen concentration (whether hypoxia or hyperoxia, Rodriguez 2008), moisture level (Sharman 2003), arterial flow (ischemia), temperature and temperature gradients (metabolic rate changes with 10%/$^o$C (White 2006, Gillooly 2001), single cells or sheets of cells moving, wound shape, shallow or deep, lean/obese, healthy/diabetic and the presence of endogenous or exogenous electric fields. The next level of factors are of the kind how macrophages and lymphocytes modulate fibroblast metabolism, protein breakdown due to the trauma, inflammatory response (Martin 2005, Medzhitov 2008) , apoptosis where material can be reused (Ewig 2008, Greenhalgh 1998), necrosis where material is completely broken down, Warburg effect (Vander Heiden 2009) and other metabolic pathways (Rabinowitz 2010). They are all characterized by being described in terms of detailed knowledge on a microscopic level. We can only account for them in case they can be understood how they affect the local metabolic rate. Furthermore there is a third level; different treatment methods including drugs, age and gender, acute versus chronic wounds, health and social staus, whether married or not, how one handles stress which influences inflammatory response, mood, mumbojumbo and gobbledygook for which we do not provide any answers in our modelling. In the end of course all this calls for a system approach. However we think that using realistic parameters in our model we can include several of these aspects in the over-all scenario.

## An energy and dimensional perspective

Let us view our own body from an energy perspective where similar balance sheets can be made for other living systems. The major part of our bodies' energy expenditure goes into the basal metabolism (at rest and no digestion) to keep up circulation, respiration, signalling, temperature and compensate for mass and heat losses. The liver, brain, heart and kidneys make up for almost 70% of this basic metabolic rate $W$ but only about 6% of the body mass $M$. To calculate gross numbers we can use a food intake equivalent to *6* MJ/day (Mahan 2000). This corresponds to an average power of *70* W or *1* mW/g = *1* pW/ng of body weight. A typical cell is around a ng in weight giving the pW scale for cell metabolism. These are quite impressive numbers in terms of human output as compared to the heat and light generated by a light bulk of *60* W. In Table I we display measured levels of cell output. Why there are major deviations from the pW regime (larger or smaller) could be an interesting research questions in itself.

**TABLE 1.** Illustrative compilation of whole cell thermal outputs.

| Microorganisms | pW/cell |
|---|---|
| Bacteria | 0.8 -5.0 |
| Yeast | 6.0-6.5 |
| Algae | 1.0-8.0 |

| Cells from Tissues | pW/cell |
|---|---|
| Fibroblasts (human skin) | 20-40 |
| Human Epidermal Keratinocytes | 21.4-112 |
| Adipocytes | 26-50 |
| Erythrocytes | 0.01 |
| Platelets | 0.06 |
| Lymphocytes | 2.0 |
| Granulocytes | 4.0 |
| Mast Cells (rat) | 5.1-20.4 |

| Intact Tissue Slices | (mW/g) |
|---|---|
| Adipose/Fat Tissue | 0.04-0.133 |
| Human Epidermis | 2-11 |

In physical terms wounds need substantial energy resources when processing the new material needed to fill the wound cavity. In the first instance it takes a lot of extra energy to boost the immune system to beat possible inflammations and clear up the debris. The energy needed for the tissue ingrowth can be decomposed into two parts. The energy stored in the new material and structures being built and the one used to maintain old and new tissue. The first one dominates during the tissue growth process and depends naturally on tissue composition. However if we know the fractions of fat and proteins one can use their heat of combustion to calculate a number for the tissue concerned. This works out to a mean energy requirement for growth of the order of *20* kJ per gram of tissue deposited or *60* μJ for an average cell (*3-5* ng) (Malina 2004). This corresponds to a power of *6* W if converted to *one* hour. Being of the order of 10% of the total basal power we see that wound healing has to take days if not to be a too heavy load on the system. Another way is to boost the local metabolism where it is well-known that in periods of illness or injury we need to increase the energy intake with up to *40-50* %. This means in physical terms that wounds acts like substantial energy sinks.

To continue we now make a dimensional analysis. Consider the body metabolism $W$ ($[W]=kg m^2/s^3$) to be decisive for the energy intake, storage and usage; being the sum of all the processes performed by cells to keep us alive. Matching the fundamental units *kilogram*, *meter* and *second* in $W$ by introducing a mass $M$ (missing mass in the wound), a length $L$ (typical size of the wound) and a time-scale $T$ (some measure of the healing time) we can form a dimensionless combination $WT^3/ML^2$ to characterize the wound healing process. Since $WT$ is the total energy $E_{tot}$ delivered and $L/T$ can be interpreted as a typical wound closing velocity *V* our dimensionless combination becomes $E_{tot}/MV^2$. Introducing the number of cells *N* and their average mass $m_c$ we can write our expression as *($E_{tot}$/N)/($m_c V^2$)* i.e. energy input per cell divided by the cell's kinetic energy. However with realistic values (*3-5* ng) for the cell mass and a typical velocity of tens of μm/hour this is way smaller than the *60* μJ ($E_{tot}$ /N) cited above as requirement for growth. Thus we need to find another variable to compare the deposited energy with.

What we have found is of course the obvious fact that the metabolic power is geared for something more than just moving the cells. In other words we need to compare $E_{tot}$ with a characteristic energy scale which is rather the cell potential energy. Taking into account that fibroblast cells are one of the major players in making new tissue we find from **Table I** and the energy requirement above of *20kJ/g* to make new tissue that a time of the order of *10* days will have to elapse to provide the necessary energy *WT*. This is definitely in line with measured healing times. Notice however that this time is basically much longer than the typical cell-doubling time or the time it takes a fibroblast cell to move 10 times its own size (*10* h), which is typically the distance over which cells have no idea that a wound is present in their neighbourhood. It is one of the virtues of our model derived below that it explains why the external and internal time-scales differ so substantially.

**Growth models**

An important aspect of metabolic resesrch is to understand and model the allocation of metabolic energy in sustenance and growth of an individual or for that matter in our case a wound. One of the most well-known models in metabolic modelling is that of von Bertalanffy (von Bertalanffy 1957, Lorenzen 1996). Let *M(t)* be the body weight at time *t*. Von Bertalanffy argued that the change in *M* with time is proportional to the difference in anabolic rate *A* (biomass synthesis) and catabolic rate *C* (biomass breakdown to provide energy and components for the anabolic process) according to

$$\frac{dM(t)}{dt} = AM^\alpha - CM^\beta \quad , \qquad (1)$$

where $\alpha$ and $\beta$ are exponents which were chosen as *2/3* and *1* respectively based on presumed surface area *S* and volume *V* scalings. We note in passing that this is in line with *simple* physical arguments where $S \propto V^{2/3}$ but differs with the measured scaling for human beings, where *S* is closer to $V^{3/4}$ [Bailey 1996].

The same equation with the exponents $\alpha = 3/4$ and $\beta = 1$ was recently introduced by West et al. (West 1999) to describe the growth of organisms. However in this case the first term is the metabolic rate of the organism ($\propto M^{3/4}$) and the second term is the sustenance of the cells created. There is therefore an important distinction with respect to von Bertalanffy's approach in that the equation is derived from first principles (energy conservation) rather than formulated *ad hoc* and that the ¾-exponent comes from a detailed model for nutrient transport in the organism. The equation as such has a very important outcome since it generates a unique growth curve where the weight development of animals from fishes, over birds to large mammals all can be brought back to the same general growth curve (West 1999). This is the more impressive if one realizes that this corresponds to mass differences of six orders of magnitude and life-spans which also differ by many orders of magnitude. We have therefore reasons to believe that there is a unique almost universal way growth proceeds in the animal kingdom. Possibly this is also true for plants and other genera.

Obviously wound healing is also a growth process, albeit taking place in the presence of the rest of the individual. In the following sections we will expand on this analogy to derive expressions for the wound healing rate in terms of an energy consideration argument.

### Energy transformed to mass

Our starting point is that the ultimate fate of all our processes in the body is heat, structures built or matter leaving the system which guides us in setting up our model. In this context there are several options of describing the wound healing process basing it on e.g. the macrophages orchestrating the whole process or the fibroblasts making the visible ingrowth of new material. We choose however to look at the possible amount of material created in the wound bed accounting for the different ways energy is used in th process. Therefore we need to address how to express the metabolic rate in terms of the mass of new cells $m(t)$, since at the end of the day wound healing aims at filling in the missing matter. For our purposes we will assume that the general relationship between body metabolic rate and body mass is also valid on a local level using a generic allometric expression

$$W = bM^\gamma. \qquad (2)$$

The dimension of the metabolic rate *W* is J/s and $\gamma$ is a scaling exponent. The pre-factor *b* is a gross number for all the microscopic biochemical reactions taking place in the pertinent cells and is thus expected to depend on all the different factors we listed in the introductory section. It is important to point out that the values of *b* and γ are not necessarily those of the whole organism where in the animal growth model of West et al. (West 2002) typical values are $\gamma = 3/4$ (slightly less for smaller animals) and *b* = *19*mW/g$^{3/4}$ (or *3.4*W/kg$^{3/4}$).

There are several justifications for using an equation of the form given in Equation (2). The first is that it captures the essential mechanisms for animal growth being empirically tested for a vast range of organisms as reviewed in Wiebel (Weibel 2002) and White (White 2005). The most appealing value for $\gamma$ from a physical point of view would be the suggestion by Rubner (Rubner 1883) that $\gamma = 2/3$ since this corresponds to surface area where eventually almost all of the energy input is going to go away as heat. However focusing on oxygen consumption instead of temperature regulation Kleiber (Kleiber 1932, 1947) claimed $\gamma = 3/4$. This $\gamma$ was recently corroborated in a theoretical analysis on a space-filling fractal nutrient exchange network for oxygen transport by West et al. (West 1999). The second argument is that we seek a simple metabolic power relationship $W = W(M)$ which vanishes for small *M* and should approach the total basal metabolic rate for the "animal" when *M* is large. Equation (2) provides such a functional relationship and can then be justified if a good correlation, with predictive power, is obtained from comparison to realistic wound healing data. Third, it conforms to the limits seen in nature for the range of masses life can sustain. For $\gamma < 1$, the specific metabolic rate $W/M \propto M^{\gamma-1}$ increases dramatically the smaller a creature becomes, finally reaching the maximum cellular metabolic rate for individual cells *in vitro*. In the other end we notice that the number of cells to support $(\propto M)$ grows faster than the rate at which they can be supplied so growth will eventually stop by itself.

We will now derive an equation, from first principle energy considerations, which provides the development of wound mass as a function of time and hence generates the wound healing rate as an outcome. We can write for the total energy input to the system $\Delta E_{tot}$ during a time interval $\Delta t$ as:

$$\Delta E_{tot} = \Delta E_{mnt} + \Delta E_{new} + \Delta E_{else}, \qquad (3)$$

where $\Delta E_{mnt}$ is the energy needed for maintenance of existing tissue at time *t*, $\Delta E_{new}$ the energy devoted to make new tissue (once made this tissue comes into the maintenance category) and $\Delta E_{else}$ is whatever other sink/source of energy in the problem which cannot be classified as new or maintenance. It can e.g. be energy used by neutrophils and macrophages in the cleaning-up part of the process or the external introduction of wound healing stimuli. It can also account for if extra energy is needed in order to compensate for unusual heat losses to the surroundings. However in general we view the final destination of the maintenance part to end up as heat and we assume the heat loss to the surroundings to be about the same for the wound area from start and when filled with new tissue; especially for true shallow wounds. Thus for the moment we will set $\Delta E_{else}$ to zero, however accounting for some of its contributions when they can be included in the other two terms in Eqs.(3).

Notice that the energy $\Delta E_{new}$ is related to the introduction of new cells in the wound area. Its main contribution is not from cell division rather it is the energy associated with tissue moving in to close a wound either in the form of a purse string mechanism (Rodriguez-Diaz 2008) for embryonic organisms and many animals or as a collective motion of cells dragging themselves over the wound matrix in the adult organism (Jacinto 2001). In the end of course, if matter is to be conserved, there has to be cell division taking place around the wound area to produce the missing biological matter. An estimate for $\Delta E_{tot}$ is now:

$$\Delta E_{tot} = W \Delta t \quad , \tag{4a}$$

where *W* is the metabolic rate which we will describe according to Equation (2) as an average local metabolic rate for the organism. Local, because one might of course suspect that the local metabolic rate is higher in a wound than in the surrounding tissue since it is a complicated machinery for tissue engineering that is launched in connection with a wound. At the same time the disturbed region can be imagined (at least in the beginning) to not function fully, e.g. with respect to nutrient transport as compared to a "bulk" situation.

We assume that at time *t* we have $n(t)$ new cells in the wound. This is of course a gross simplification but staying at this more macroscopic level we imagine the wound being filled with a generic cell type having the necessary characteristics as we will dwell on below. We can then write the maintenance cost as

$$\Delta E_{mnt} = n(t) W_c \Delta t \tag{4b}$$

where $W_c$ is the metabolic rate of a single cell *in vivo*. Likewise if $E_c$ is the energy needed to assemble a given cell/move an existing cell into the wound volume

$$\Delta E_{new} = [\frac{d}{dt}(n(t) + n_d(t))] \Delta t \cdot E_c \quad . \tag{4c}$$

The second term in Equation (4c) is to account for the loss of cells by different causes ($n_d$). A reasonable assumption would be that $dn_d / dt = n / T_d$ where $T_d$ is a time constant for the viable cells to die or be removed in some other way. Whether it is necrosis, autophagy (prone in nutrient poor environments) or apoptosis we envisage that the different processes add in proportion to $1/T_d$ as a sum of their inverse time-constants. We can thus include this term, which is now proportional to *n(t)*, in a renormalized cellular metabolic rate $\overline{W_c} = W_c + E_c / T_d \equiv W_c(1 + T/T_d) = E_c(1/T + 1/T_d)$, in Eq.(4b), introducing a time-scale for assembling a cell $T = E_c / W_c$ for later use. Notice that the renormalized metabolic rate increases considerably in proportion to $T/T_d$. There is also a renormalization of $E_c$ (denoted $\bar{E}_c$) for the following reason. We are using a description where we consider an average or generic wound filling cell. In this respect one will have to include in $E_c$ not only the energy needed in assembling/moving a cell but also its share of the surrounding extra-cellular matrix. We are thus talking of cells in the model dressed up with their surrounding medium as compared to the "free" cells. This also means that the cellular mass $m_c$ is renormalized to $\bar{m}_c$, and in this manner also any change in cell volume will be included in our estimates. In all this also gives a new time-scale of the problem which is $\bar{T} \equiv \bar{E}_c / \bar{W}_c = TT_d /(T+T_d)$. Notice that the "death" of cells show up in the characteristic time-scale of our wound healing scenario implying that a measurement of a real wound healing time trajectory includes both generation as well as loss of cells.

Combining Equations (2-4) we can now write:

$$W = n(t)\bar{W}_c + \bar{E}_c \frac{dn(t)}{dt}. \tag{5}$$

From Equation (2) we have $W = bM^\gamma$ and writing the total mass $M(t) = n(t)\bar{m}_c$ in terms of $\bar{m}_c$ Eq.(5) now takes the form

$$(\bar{E}_c / \bar{W}_c)\frac{dM(t)}{dt} = (\bar{m}_c b / \bar{W}_c)M^\gamma(t) - M(t). \tag{6}$$

Being first order in time derivative we need to specify one boundary condition. We choose this as the starting mass $M(t=0) \equiv M_o$ and will comment upon this choice further down.

In order to illustrate the main aspects of Eq.(6) we will write it on dimensionless form. $\bar{E}_c / \bar{W}_c$ has the dimension of time and is the average time $\bar{T}$ it takes to make a new dressed cell (at the cost $\bar{E}_c$) given the available energy input $\bar{W}_c$ per unit time. Appendix A discusses the major factors contributing to $\bar{E}_c$ (i.e. $\bar{T}$). For the mass scale, we use the pre-factor of the first term on the right hand side of Eq. (6). It is related to the maximum possible mass attainable in the process at which time all incoming energy is used for maintenance of the organism, if it is not dead before that. In principle this means that we can have a situation that the wound cavity cannot be filled $M_\infty \leq M_w$, where $M_w$ is the mass necessary to fill the wound cavity. The opposite situation can of course also be the case $M_\infty \geq M_w$. From the stationary solution to Eq.(6) we have

$$M(t \to \infty) \equiv M_\infty = (\bar{m}_c b / \bar{W}_c)^\alpha, \tag{7}$$

where $\alpha \equiv 1/(1-\gamma)$ is a number directly given in terms of $\gamma$. It will turn out to play an important and decisive role in what follows since it is expected to be in the range of 2-6 thus making $M_\infty$ a very strong function of $\overline{m}_c$, $b$ and $\overline{W}_c$.

Measuring the mass $M(t)$ in units of $M_\infty$ ($=\overline{m}$) and the time $t$ in units of $\alpha T$ ($=\bar{t}$) we can from Equation (6) write down the following equation for the development of the cell mass in a wound:

$$\frac{d\overline{m}(\bar{t})}{d\bar{t}} = \alpha\left(\overline{m}^\gamma(\bar{t}) - \overline{m}(\bar{t})\right). \tag{8}$$

This generic growth equation has the explicit solution

$$\overline{m}(\bar{t}) = [1 - (1 - \overline{m}_o^{1/\alpha})e^{-\bar{t}}]^\alpha, \tag{9}$$

where $\overline{m}_o \equiv \overline{m}(\bar{t}=0)$ is the starting mass in units of the final mass attainable, and it is the expression describing the development of the new tissue mass in the our wound model. The actual wound mass $M_w(t)$ is assumed proportional to $\overline{m}(\bar{t})$. Notice however the difference between the possible attainable mass $M_\infty$ in Eq.(7) and the mass actually needed to complete the wound healing process $M_w(t)$.

**Growth versus maintenance**

We can gain further insights from our model by looking at the energy expenditure in terms of the part of the energy which goes to *maintenance* as compared to *total* energy input

$$\frac{\Delta E_{mnt}}{\Delta E_{tot}}(\bar{t}) \equiv \frac{n(t)W_c}{W} = \overline{m}^{1/\alpha}(t) = 1 - (1 - \overline{m}_o^{1/\alpha})e^{-\bar{t}}. \tag{10}$$

This maintenance cost increases linearly with time $\bar{t}$ in the beginning (starting from $\overline{m}_o^{1/\alpha}$) and exponentially approaches the situation where it totally dominates the energy expenditure. The corresponding fraction of energy going to growth is by definition:

$$\frac{\Delta E_{new}}{\Delta E_{tot}}(\bar{t}) \equiv 1 - \frac{\Delta E_{mnt}}{\Delta E_{tot}}(\bar{t}) = (1 - \overline{m}_o^{1/\alpha})e^{-\bar{t}}. \tag{11}$$

Starting out using a fraction $1 - \overline{m}_o^{1/\alpha}$ of all available energy and rapidly approaching zero. Equations (10) and (11) make it possible for us to write the tissue mass increase in the more suggestive form

$$\overline{m}(\overline{t}) = [1 - \frac{\Delta E_{new}}{\Delta E_{tot}}(t)]^\alpha = [\frac{\Delta E_{mnt}}{\Delta E_{tot}}(t)]^\alpha. \tag{12}$$

The discussion above shows that growth/maintenance makes up for everything/nothing in the beginning but is a very small/large fraction at the end of the growth process. These considerations could be of importance when designing a wound healing treatment since different agents, such as specific growth factors, will have to be administrated during different temporal parts of the wound trajectory (Robson 2000 & 2001). Since α is typically in the range *2-6* the dependence in Eq.(12) is a very strong function of the relative energy utilization.

A very interesting aspect of our model, which should have relevance when monitoring wound trajectories, is that the growth has to peak somewhere in between the empty wound bed and the completed growth process. We therefore form the second derivative of $\overline{m}(t)$:

$$\frac{d^2\overline{m}(\overline{t})}{d\overline{t}^2} = -\alpha r_\alpha e^{-\overline{t}}(1 - \alpha r_\alpha e^{-\overline{t}})(1 - r_\alpha e^{-\overline{t}})^{\alpha-2} \tag{13}$$

where $r_\alpha \equiv 1 - \overline{m}_o^{1/\alpha}$. Eq.(13) vanishes for $\overline{t} = \ln r_\alpha$, $\overline{t} = \infty$ and $\overline{t} \equiv \overline{t}_{max} = \ln(\alpha r_\alpha)$. The first one is outside the time interval to the left (approaching $\overline{t} = 0$ when $\overline{m}_o \to 0$) and the second one corresponds to the final situation as $\overline{t} \to \infty$. The third one corresponds to the <u>inflection point</u> in the growth curve (which exists if the starting mass $\overline{m}_o \leq (1 - 1/\alpha)^\alpha \equiv \overline{m}(\overline{t}_{max})$, the mass at the inflection point) with a maximum in the wound healing rate

$$\frac{d\overline{m}(\overline{t} = \overline{t}_{max})}{d\overline{t}} = (1 - \frac{1}{\alpha})^{\alpha-1} \tag{14a}$$

and the corresponding mass

$$\overline{m}(\bar{t}_{\max}) = (1 - \frac{1}{\alpha})^{\alpha} \leq \frac{1}{e}, \qquad (14b)$$

where the last line corresponds to the limit of infinite α-values (γ approaching 1). When the starting mass $\overline{m}_o \geq \overline{m}(\bar{t}_{\max})$, the mass at the inflection point, the wound mass increases in a monotonic fashion between $\overline{m}_o$ and 1.

Notice that both values in Eq.(14) are independent of the original mass (though the exact timing is) only depending on the scaling coefficient in the metabolic rate. Hence the peak value of the wound healing rate has direct information about the scaling of the local metabolic rate. Moroever it is a strong function of $\gamma$ since it enters in the form of the construction $\alpha \equiv 1/(1-\gamma)$. The very small *12.5 %* increase when going from $\gamma$ being *2/3* to *3/4* , γ values which has created fierce debates in the litterature, will in the wound healing scenario correspond to an almost three times larger difference between the two.

Before studying the details of the model we have to set the general growth scenario into the wound healing perspective, i.e. the relationship between the needed wound mass and the maximum attainable mass $M_{\infty}$ in the growth proces. All masses are measured in terms of the latter. We compare the starting mass $\overline{m}_o$, the maximum growth rate mass $\overline{m}(\bar{t}_{\max})$ and the mass needed to fill the wound cavity $\overline{m}_w$. We consider two general cases. *A)* If $\overline{m}_w \geq 1$ the situation is such that the growth is not able to fill the wound cavity and *B)* if $\overline{m}_w \leq 1$ the situation is such that the growth is able to fill the wound cavity.

The first situation is an akward one from the point of view of wound treatment and is directly correlated with the factors entering Eq.(7) for total possible mass. The pre-factor *b* in the metabolic rate enters explicitly by setting the mass-scale $M_{\infty} \propto b^{\alpha}$. This in itself indicates different strategies if one wants to influence the final wound mass. For typical values of γ, α is in the range *2-6* and therefore small changes in $\overline{m}_c$, *b* or $\overline{W}_c$ can have dramatic effects on possible final wound mass. We would on general grounds believe e.g. that the prefactor *b* in Equation (2) is smaller for a diabetic wound than a normal wound hence influencing the wound characteristics in a strong way. We also want to point out that b has been found to be a strong function of temperature (Gillooly 2001).

In the second case we have mass enough to fill the wound cavity however there are two possible (extreme) scenarios. One is that we start from zero mass ($\overline{m}_o = 0$) and build new tissue, thus $\overline{m}_w >> \overline{m}_o$. This situation is most reminiscent of individual cells and cell sheets crawling into the wound bed. If furthermore $\overline{m}_w > \overline{m}(\bar{t}_{\max})$ there is then the possibility of seeing effects of the maximum growth rate; there should be a visible peak value in the rate of wound healing.

The other extreme situation is if the wound healing process is seen as the body is just filling in the missing mass to complete the total mass. In this case $\bar{m}_o = (M_\infty - M_w)/M_\infty \equiv 1 - \bar{m}_w, \bar{m}_w \ll 1$. This means functional tissue is added to the wound area as a growth on the latter stage of the general growth curve. We are then definitely in a mass interval where there is no influence of the inflection point in the growth curve. This situation is most reminiscent of existing cells moving into the wound bed through a mechanism parallel to the wound perimeter: the so called purse-string mechanism (McGrath 1983) as for rats, rabbits and human fetuses. From Eq.(9) we find that in this case:

$$\bar{m}(\bar{t}) \approx 1 - \alpha r_\alpha e^{-\bar{t}} \quad . \tag{15a}$$

In other words

$$\ln[1 - \bar{m}(\bar{t})] \approx \bar{t}_{\inf} - \bar{t} \, , \tag{15b}$$

where $\bar{t}_{\inf} \equiv \ln \alpha r_\alpha \approx \ln \bar{m}_w$ is negative and outside of the positive time interval. For processes where the wound closes in order to complete the full body growth process we thus expect experimental results which are almost strictly exponential. Also the extrapolated values to small times reflect the value of $\bar{m}_w$ yielding an estimate of the possible maximum mass attainable given we know the amount of missing mass corresponding to the wound volume.

**Model predictions**

For all the different cases considered above we can use Eq.(9) to find the actual wound mass growth curve from the generic one; $\bar{m}(\vec{t})$. In what follows we therefore give a representation of the main features of our wound healing model by looking at $\bar{m}(\bar{t}) = [1 - e^{-\bar{t}}]^\alpha$, i.e. for $\bar{m}_o = 0$, using five representative values of $\gamma$ (*1/2, 2/3, 3/4, 4/5* and *5/6* respectively). This corresponds to values for $\alpha = $ *2, 3, 4, 5* and *6,* and the corresponding $\bar{m}(\vec{t})$ are shown in Figure 1. These values of α means that if the normalized time $\bar{T}$ is interpreted as an inherent effective cell-division time, or the time to get a new cell into the wound, including the assembly of concomitant surrounding extra-cellular matrix, the *apparent* time scale in wound healing measurements $\alpha \bar{T}$ is 2-6 times longer than $\bar{T}$ ~ *1* day. If we were to plot instead $\ln(1 - \bar{m}(\bar{t})^{1/\alpha})$ this would be directly a straight line $-\bar{t}$ and we can easily extract $\alpha \bar{T}$.

For short times $\bar{t} < 1$, $\bar{m}(\bar{t})$ goes as $\bar{t}^\alpha \sim \bar{t}^{2-6}$ using the typical values above for the metabolic rate scaling exponent γ. Hence the matter growth in the begining increases extremely slowly with time (Figure 1). The implication of this is that it will take about one time unit before any visible signs of tissue growth will be apparent. Often fits to real wound healing data (Cukjati 2001) is done only in the part which has a visible variation claiming that there is an initial phase in which the healing process is not yet started. Our claim is that with the growth varying as $\bar{t}^\alpha \sim \bar{t}^{2-6}$ for small $\bar{t}$, there is no way of experimentally claiming a delayed wound healing based on the wound area measurements alone. For large $\bar{t}$ $\bar{m}(\bar{t})$ approaches 1 exponentially fast $(1 - \alpha e^{-\bar{t}})$.

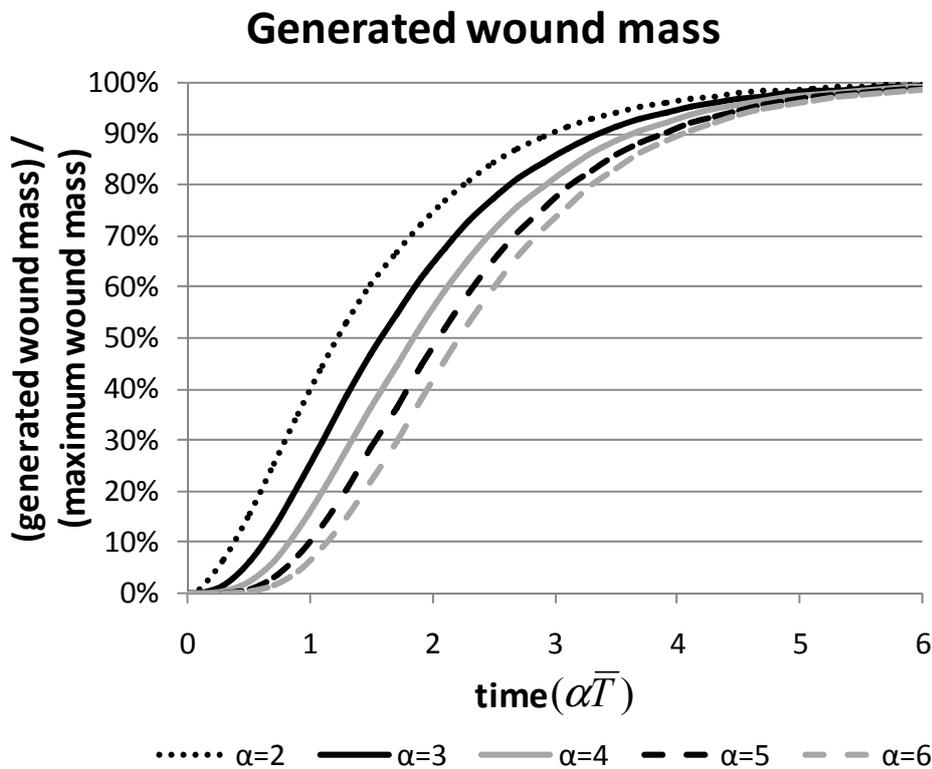

**Figure 1.** The wound tissue mass increase is shown as a function of time. Mass is measured in units of the maximum possible mass which can be generated, starting from no mass, and time is measured in units of $\alpha \bar{T}$ where $\bar{T}$ is an intrinsic timescale for the wound measuring the typical time for generating new cells within the wound area and $\alpha \equiv 1/(1-\gamma)$. γ is the exponent relating mass to metabolic rate. γ=2/3 corresponds to a classical thermodynamic scaling due to heat losses and γ= 3/4 corresponds to a nutrient transport network model. As can be seen the curves have a very similar appearance starting out very slowly in the beginning, going through a rapid change when 1 < time < 3 and again entering a slower phase for longer times. Notice that for the range of γ-values shown α varies from 2 to 6 thus having a drastic influence on the final apparent time-scale compared to the intrinsic time-scale $\bar{T}$ for the system. Also notice that the curves above correspond to maximum attainable mass and this should eventually be compared to the mass which actually is needed to fill the wound cavity.

In Figure 2 we plot the wound healing rate $d\bar{m}/d\bar{t}$ as a function of normalized time $\bar{t}$. Numerically the maximum wound healing rate is *4/9 (mass 8/27)* for *α=3 (γ=2/3)* and *27/64 (mass 81/256)* for *α=4 (γ=3/4)* approaching *1/e* from above (mass from below) for large values of $\alpha$. Whereas the peak value of the wound healing rate is rather insensitive to the value of $\alpha$ being of the order of *0.4*, the time instant when it peaks is a strong function of α. In this way the peaking of the wound healing rate can be a way of discriminating $\gamma$-values differing by a mere 8% (2/3 and 3/4 respectively) since this corresponds to a 70% difference in the time (inflection time $\propto (\alpha \ln \alpha)\bar{T}$) when it happens. We suggest therefore that the measured healing rate could also be used in assessing the wound healing trajectory; if it progresses the way it should or not.

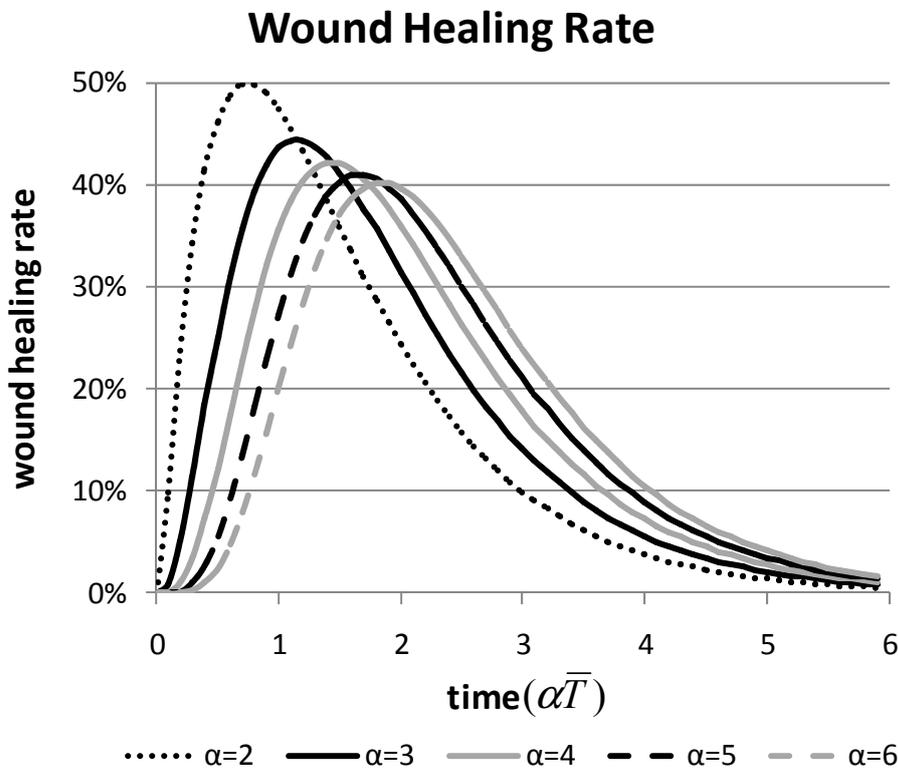

**Figure 2.** Wound healing **rate** as a function of time. Mass is measured in units of maximum attainable mass and time is measured in units of $\alpha\bar{T}$ where $\bar{T}$ is an intrinsic timescale for the wound, measuring the typical time for generating new cells within the wound area. $\alpha \equiv 1/(1-\gamma)$ is a number indicated in the figure and is related to $\gamma$, the mass-scaling coefficient of the metabolic rate as described in the legend to Figure 1. The wound healing rate peaks very quickly for times $\bar{t}$ of the order of $\ln\alpha$ after a slow start and a similarly slow final part. All curves have by definition the same area ($=\bar{m}(\infty)-\bar{m}(0)\equiv 1$).

In order to connect our model with the typical plot in wound care management, which is usually made in terms of montoring remaining open wound area *A(t)* versus healing time *t* (Gilman 2004), we need relate our $\overline{m}(\bar{t})$ with the experimental measurements of *A(t)*. Typically the latter is normalized to the original area *A(t=0)* thus starting out at unity and ending up at zero (*A(t)/A(t=0)*). Especially for the shallow wounds we are considering all the generated mass should be directly linked to the covered area so we expect *A(t)/A(t=0)* = $1 - \overline{m}(\bar{t})/(1-\vec{m}_o)$.

In the shallow wound situation described above we then expect from Eq.(15b) that

ln[ *A(t)/A(t=0)*] = $-\bar{t}$ . (16)

Plotting the logarithm of how the normalized wound area changes with time we expect a straight line which gives us direct information on the effectice time-scale of the problem. We therefore now turn to the experimental part of our study.

**Experimental Procedures**

Animal procedures were conducted in accordance with the Guide for the Humane Care and Use of Laboratory Animals. The experimental protocol was approved by Drexel University's Institutional Animal Care and Use Committee (IACUC).

*Hairless Rats:* The results of a study describing the optical properties of wounds in these hairless rats was published previously [Papazoglou 2008]; here we report only the measured wound sizes from these animals as they relate to our model of healing. Nine Sprague-Dawley female hairless rats were purchased from Charles River Laboratory (Wilmington, MA). The rats were ten weeks of age and weighed approximately 205 g at the time of diabetes induction. Four rats were rendered diabetic using intraperitoneal injection of streptozotocin (STZ) at 75 mg/kg. In order to assure successful induction of diabetes, blood glucose levels were monitored weekly, and remained greater than 200 mg/dl in injected (diabetic) rats and less than 200mg/dl in control (healthy) rats throughout the study. Wound surgery was performed nine days after diabetes induction, as described below.

*Zucker Diabetic Fatty Rats:* Five female obese and three female lean Zucker Diabetic Fatty (ZDF) rats were purchased from Charles River Laboratories (Wilmington, MA). The rats were 21 weeks old at the start of the study, and the mean weights of lean and obese rats at the time of surgery were 197 g and 410 g, respectively. Blood glucose levels in all animals were monitored weekly and remained less than 200 mg/dl throughout the duration of the study.

*Animal Handling Procedures*: Throughout the course of the study animals were housed in individual cages on alpha cellulose bedding and maintained in an animal care facility with a 12 hour light and dark cycle. Food and water were supplied *ad libitum*. On the day of surgery, one full thickness wound of 4.6 cm$^2$ was made on the left side of the dorsal area of each animal using sterile technique in an animal surgical suite. The surgery was performed using isoflurane anesthesia administered via a face mask. All wounds were covered with a Tegaderm sterile transparent dressing (3M, Minneapolis, MN) after wound surgery and between wound measurements.

*Wound Size Measurement;* Wounds were digitally photographed using a Fujifilm Finepix s700 digital camera 2-3 times per week with cross-polarizing filters to reduce surface reflection. A ruler was held in the imaging plane of each photograph to allow the calculation of absolute wound area. The boundaries of each wound were manually traced using a computer mouse and Microsoft Paint software, and then wound areas were calculated from the traced photographs using Matlab (Mathworks, Inc.) software image analysis tools. The image analysis procedure counts the number of pixels within the traced wound boundary, and then calculates the wound area by approximating the size of each pixel from the image of the ruler that was present in the imaging plane of each wound.

*Results:* Normalized wound area $A_{norm} = A(t)/A(t=0)$ was obtained for each wound on each measurement day by calculating the ratio of wound area each day to the initial wound area on the day of surgery (day 0). Average normalized wound areas on each measurement day are presented in Figure 3. The obese ZDF wounds remained open for 33 days after wound surgery, lean ZDF wounds remained open for 22 days, while both hairless STZ-induced diabetic wounds and hairless non-diabetic wounds remained open for 21 days.

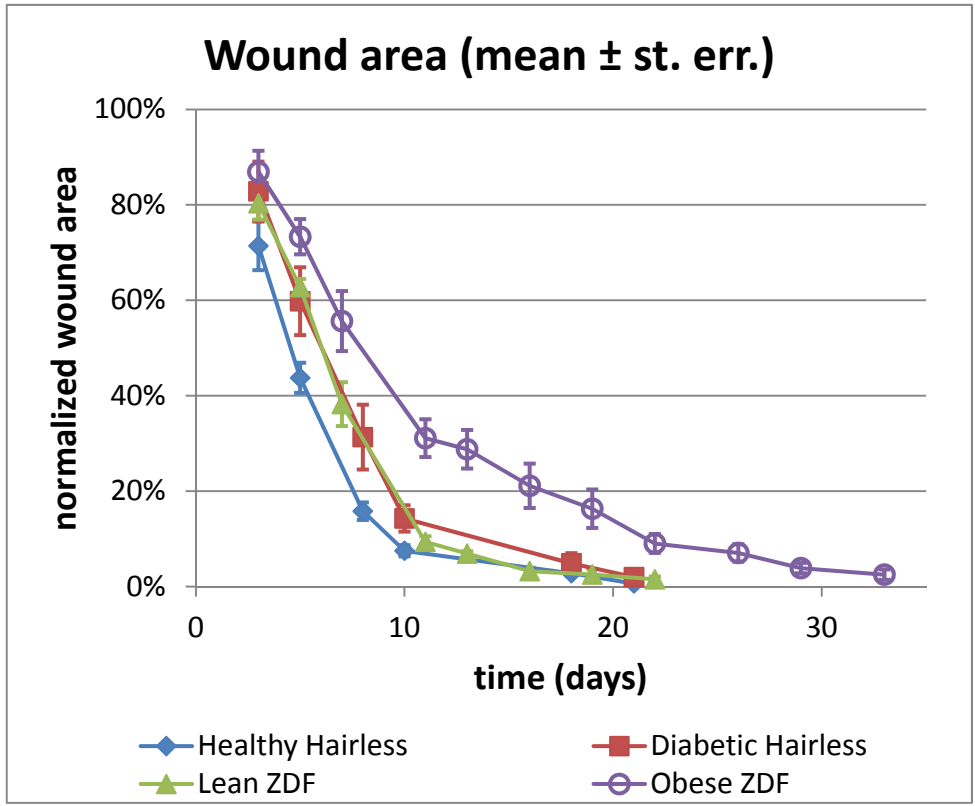

**Figure 3**: Normalized wound areas (mean ± s.e.) from obese ZDF (rings), lean ZDF (triangles), non-diabetic (healthy) hairless (diamonds), and STZ-induced diabetic hairless rats (squares) are compared. We notice that the obese ZDF takes much longer time to heal while the other three are very much similar with respect to full healing. Healthy hairless is quickest to achieve a certain degree of healing, however merging with lean ZDF after some time.

We find that changes in the surface area of full-thickness excisional wounds can to a good degree be represented with an exponential model, returning below to deviations from this. The normalized wound areas measured on each animal that survived until the end of each study were fitted to the equation $A(t)/A(0) = e^{-t/\tau}$, where *A(t)* is the wound area at time *t* (days). From the experiments we can extract the wound healing rate $1/\tau$ (% per day). We summarize in Table 2 below the results in terms of the wound healing rates and in Figure 4 we show the curves these numbers are extracted from. Notice in this context that the wound healing rate $1/\tau$ above corresponds to $1/\alpha\overline{T}$ in the model given in Eq.(9). Since for a typical metabolic scaling $\alpha \approx 4$ the experimental results correspond to $\overline{T}$-values around a day, however two days for the Obese ZDF. This two-day time-scale is thus much closer to the one expected from general considerations of cell proliferation (Hehenberger 1998) than the apparent eight-day time scale in the wound experiments. Thus our model gives a general insight why the wound time-scale differs from the intrinsic time-scales of importance in the wound healing process.

**Table 2**. Wound healing rates and time-scales for selected animals as extracted from the experimental data in Figure 3. We see the wound healing rates fall in a reasonable fashion given the expectations from the medical conditions. Taking the inverse of the healing rate we get the typical time-scale for wounding. In brackets we have divided this with a typical $\alpha \approx 4$ to exhibit the general intrinsic time-scale of the processes being responsible for the healing process.

| Animal | Wound healing rate (% per day) | Wound healing rate time-scale $\alpha\overline{T}$ (days) |
|---|---|---|
| Healthy Hairless | 24 | 4.1 [1.0] |
| Lean ZDF | 22 | 4.5 [1.1] |
| Diabetic Hairless | 20 | 5.0 [1.2] |
| Obese ZDF | 12 | 8.5 [2.1] |

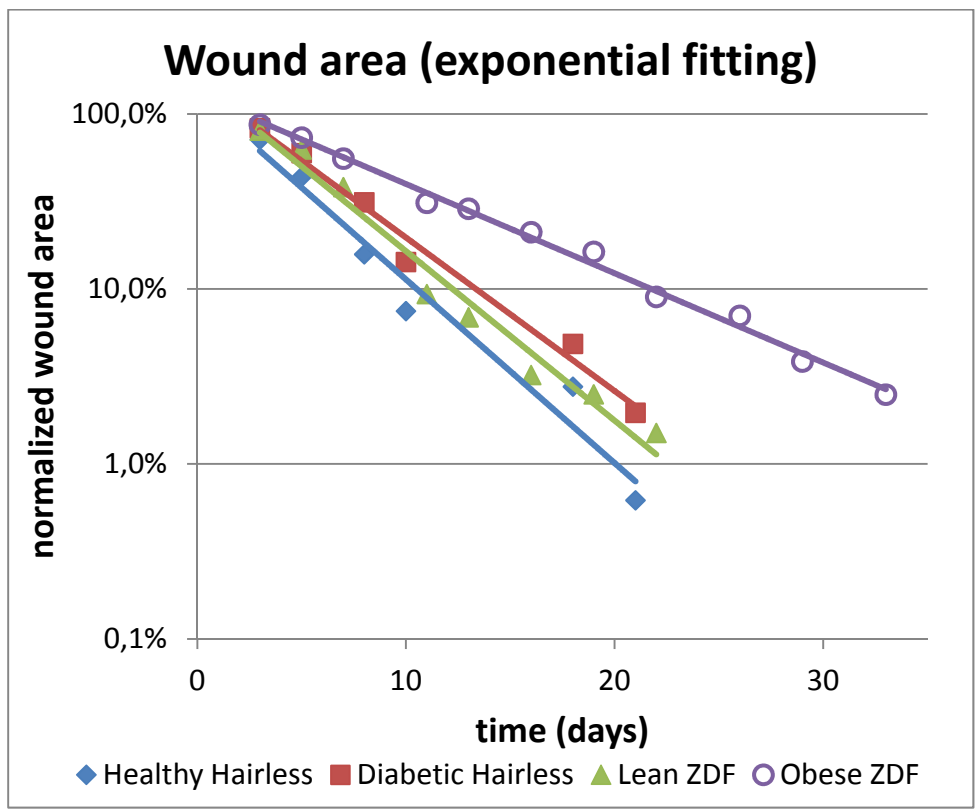

**Figure 4**: The logarithm of the normalized wound areas (mean ± s.e.) from obese ZDF (rings), lean ZDF (triangles), non-diabetic (healthy) hairless (diamonds), and STZ-induced diabetic hairless rats (squares) are compared. We notice that there is generally a good fit to a straight line and that the slopes make sense regarding the medical conditions (see also Table 2).

From the general aspects of our model and the experiments above it is clear that at the outset there is little change in the wound area covered as well as at the end. It is therefore of interest to study in more detail how the model behaves and under what conditions there is a maximum in the wound healing rate in between. In order to do this one would have to monitor the wound development on a shorter time-scale than the present one of 2-4 days in between. However to give a gross picture how it looks like we show in Figure 5 the derivative of the wound data presented in Figure 3. Whereas most of the wound healing can be described as a simple exponential (Figure 4) we see that there's a clear maximum in the rate after 5-10 days.

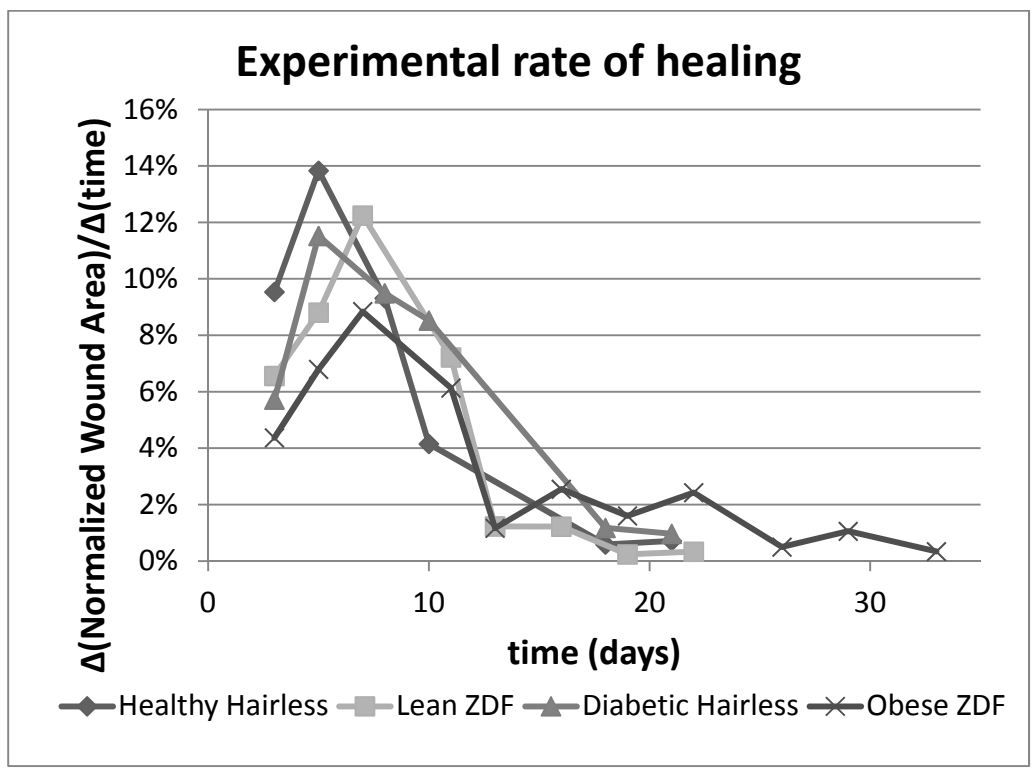

**Figure 5**. Experimental wound healing rates based on taking the derivative of the data shown in Figure 3 including the starting point at zero time. According to the figure, the peak rate of change for healthy hairless and diabetic hairless rats falls on day 5, while the peak rate of change for lean and obese ZDF rates occurs on day 7. The figure is only indicative of a peak rate since it is difficult to exactly pin-point the peaks because of 2-4 days between measurements.

## Discussion

We have in this paper presented a simple model for the growth of new tissue filling a wound cavity and compared this to experimental results on rats. We are presently applying our formalism also to analysis of human wound data.

Considering the most fundamental of the physical conservation laws we use energy conservation to connect mass development to time, which of course is crucial for a wound healing trajectory measurement. In this way we have considered only a temporal description in a highly spatially inhomogeneous situation – something we will follow up in the second paper in this series.

Even though the whole situation of wound healing is a very complex our mode of operation has been to formulate a simple model to see what predictions it gives and what constraints and connections to various physiological quantities it involves. Our model depends crucially on the scaling relationship between metabolic power $W$ and mass $M$ according to $W = bM^\gamma$ where $\gamma$ is a crucial scaling parameter. In terms of this we have made the following major observations:

- the apparent measured time-scale for wound healing should be a factor $\alpha = 1/1-\gamma$ larger than the intrinsic wound-healing time scale thus providing an answer why wound healing happens over weeks rather than days since $\alpha$ i????pi????????????????

- the maximum possible mass which can be generated in the growth process is $M_\infty = (\overline{m}_c b / \overline{W}_c)^\alpha \equiv (b/b_c)^\alpha$ expressed in terms of the basic and fundamental properties entering our model where $\overline{m}_c$ is the mass of a cell with its surrounding extracellular matrix, $b$ is the metabolic pre-coefficient and $\overline{W}_c$ is the metabolic rate for a single cell taking cell death into account. In the last line we have defined a metabolic rate for an individual cell *in vitro* $b_c = \overline{W}_c / \overline{m}_c$ and we see that the maximum attainable wound mass is the ratio between the metabolic rate pre-factor for cells in tissue as compared to the same for an isolated cell. This means that for situations where cells for different reasons prefer to be on their own rather than building tissue little wound mass is generated. Notice also the strong dependence on $\alpha$.

- the wound healing rate is maximum at a time instant, and with corresponding mass, only depending on the value of $\alpha$. One characteristic feature found is that chronic wounds notoriously show the absence of a peak in the derivate of the wound healing rate. We thus advocate that wound healing data rather should be plotted as the change in wound healing rate with time rather than just the wound healing rate. Another reason for doing this is that patients are not typically seen by clinicians immediately after a wound is inflicted, therefore wound area measurements often cannot be obtained for small $\bar{t}$, resulting in growth curves that often begin at an unknown point in the middle of the healing process.

Finally it is our hope that this model of ours can inspire to new ways of looking at wound healing as well as monitoring real wounds.

### Acknowledgement

This project was supported by Adlerbertska Foundation. I am deeply grateful for insightful discussions with Professors Colin D McCaig and Bo Ekman. The inspiration and contributions from Elisabeth S. Papazoglou have been crucial for the progess of this paper, unfortunately she has laid down her earthly tools not being with us any longer.

## Appendix A

In this appendix we give a background to the three contribtions to the intrinsic time-scale $\bar{T}$ of our wound healing model. $\bar{T} = \bar{E}_c / \bar{W}_c$ where $\bar{E}_c$ is the energy associated with producing one new cell and $\bar{W}_c$ is the metabolic rate of a single cell. The bars indicate that we are not dealing with naked cells rather cells dressed up with their surrounding medium and in the case of $\bar{W}_c$ also with cell death during the process. Below we use the unrenormalized quantities for ease of discussion.

If new cells were the source of matter for filling the wound cavity $T$ would simply be proportional to an average cell doubling time. However for the rat wounds we consider a purse string mechanism is of major concern for closing the wound and in human wounds it is the collective transport of cells into the wound area which makes up for the major part of the wound closing. In general then (neglecting cell death) we can write for $E_c$

$$E_c = pE_{cd} + (1-p)E_{pc}), \tag{A1}$$

for a fraction *p* of the available energy going into cell division (cd) and a fraction *1-p* utilizing the purse string/collective transport mechanism (pc). A simple estimate of p would be that it is proportional to the relative energy cost for that process; i.e. $p = E_{pc}/(E_{pc} + E_{cd})$. This means that we can write Equation (A1) on the form

$$E_c \approx \frac{E_{cd} E_{pc}}{E_{cd} + E_{pc}}. \tag{A2}$$

If cell doubling is the most likely process (having the smallest energy need) it gives the major contribution to $E_c$ and vice versa. The possible processes are in other words coupled in parallel and we can hence write for the time-scale *T*:

$$1/T = 1/T_{cd} + 1/T_{pc}, \tag{A3}$$

where $1/T_{cd} \equiv W_c/E_{cd}$ and $1/T_{pc} \equiv W_c/E_{pc}$. This is thus an extra complication influencing that the real time-scale in wound measurements differ from the intrinsic ones $T_{cd/pc}$. Had we also included cell-death through $\bar{W}_c$ Eq.(A3) would have had the added term $1/\tau_d$.

**Appendix B**

In this Appendix we adress the point that the crucial factor α governing the wound healing scenario diverges for γ=1 and we thus need to treat this point separately. For this situation we can write Eq.(6) in the following way

$$\frac{dM(\bar{t})}{dt} = (S-1)M(\bar{t}) \tag{B1}$$

introducing $S = \overline{m}_c b / \overline{W}_c \equiv b/b_c$ as a dimensionless measure governing the time evolution. In the last line we have introduced a *specific* metabolic rate coefficient $b_c \equiv \overline{W}_c / \overline{m}_c$ for a γ=1 situation. Notice that in contrast to Eq.(8) there is no longer any intrinsic mass-scale $M_\infty$. If the cellular specific metabolic rate $b_c$ is larger than the tissue *specific* metabolic rate $S<1$ and *M* decays with time; i.e. there is no sense in making tissue if the cell itself is more effective per unit mass. We will have a situation where the wound does not heal. In the opposite situation ($b > b_c$) M grows without bound. So for γ=1 and a situation where tissue is more "effective" than the cells themselves we have a situation where our model diverges which we could tentatively be interpret as a cancer-like growth in the broadest sense of the word. We note in passing in this context that the basic model used in this paper has been applied earlier to the ontogenetic growth of multi-cellular tumor spheroids (Condat 2006).